\documentclass[aps,physrev,preprint,groupedaddress]{revtex4-2}
\usepackage[total={6in, 8in}]{geometry}
\usepackage{amssymb}
\usepackage{caption}
\usepackage[labelformat=simple]{subcaption}

\usepackage{amsmath}
\usepackage{soul}
\usepackage{xcolor}
\usepackage{multirow}
\usepackage{graphicx}

\begin{document}


\title{Electric-field effects on defect migration energetics in GaN}



\author{Farshid Reza}
\affiliation{Department of Nuclear Engineering, The Pennsylvania State University, University Park, PA 16802, USA}
\author{Hamdy Arkoub}
\affiliation{Department of Nuclear Engineering, The Pennsylvania State University, University Park, PA 16802, USA}
\author{Alexander S. Hauck}
\affiliation{Department of Nuclear Engineering, The Pennsylvania State University, University Park, PA 16802, USA}
\author{Adri C.T. van Duin}
\affiliation{Department of Mechanical Engineering, The Pennsylvania State University, University Park, PA 16802, USA}
\author{Miaomiao Jin}
\email[]{mmjin@psu.edu}
\affiliation{Department of Nuclear Engineering, The Pennsylvania State University, University Park, PA 16802, USA}

\date{\today}

\begin{abstract}
A predictive understanding of defect transport in GaN under operating electric fields is critical for assessing device reliability in high-power and radiation environments. In this work, a ReaxFF reactive force field  for GaN is developed using a density-functional-theory training set that includes structural, thermodynamic, and defect properties. The force field yields various properties such as lattice parameters, cohesive energies, and defect formation and migration energies in close agreement with prior first-principles and experimental results. Under externally applied electric fields, we find that migration barriers can be strongly modulated, with changes that depend on defect type and field orientation. Notably, the electric fields do not simply linearly bias defect motion in GaN, but can anisotropically modify migration barriers through charge-lattice coupling, leading to nonlinear transport behavior. The response arises from field-induced partial charge redistribution and local lattice distortion. These results demonstrate that electric fields can complexly modify the defect migration landscape, providing new insight into defect transport in GaN under high-field conditions.
\end{abstract}


\maketitle

\section{Introduction}
\label{intro}
III--nitride semiconductor materials, GaN and AlGaN alloys are increasingly recognized as promising candidates for photocatalytic~\cite{yoshida2010photoluminescence} and thermoelectric applications~\cite{sztein2014polarization}, attributed to their advantageous electronic properties and high thermal robustness ~\cite{dridi2002pressure, mishra2008gan, fletcher2017survey, piprek2012ultra, cciccek2024review}. In addition, AlGaN alloys are well suited for extreme environments such as spacecraft electronics, where radiation resistance is indispensable for maintaining long-term device stability and performance~\cite{pearton2015ionizing}. Irradiation can induce point defects, defect complexes, and microstructural damage that degrade charge transport characteristics and adversely affect long-term device reliability~\cite{liu1998ion, kucheyev2000ion, islam2019heavy, sequeira2021unravelling, kuboyama2011single,mahfuz2024microstructural,sequeira2021unravelling,sequeira2022examining}. In particular, radiation-induced defects can create deep-level states in the bandgap that facilitate carrier recombination and localization, leading to enhanced nonradiative losses, carrier scattering, and undesirable optical transitions~\cite{pimputkar2009prospects,jani2007design}. 

Beyond defect formation, defects in operating semiconductor devices are not static; they migrate and reconfigure under the combined influence of electric field and temperature, driving long-term performance drift. The relevance is even greater in GaN-based high-electron-mobility transistors (HEMTs) and power devices, which routinely operate under fields approaching the breakdown limit ($\sim$3~MV/cm) and at elevated junction temperatures, and additionally carry large intrinsic polarization fields at AlGaN/GaN heterointerfaces~\cite{ambacher1999two}. Under high-field operation, electrical-stress-induced trap generation and the associated threshold-voltage instabilities are well documented in these devices, with degradation kinetics that are further accelerated at elevated junction temperatures~\cite{meneghesso2008reliability, tapajna2010integrated}. Capturing how the electric field modifies defect migration energetics beyond defect generation under radiation is therefore essential for predicting long-term device behavior under realistic operating conditions.

Multiple experimental techniques such as deep-level transient spectroscopy (DLTS) \cite{kanegae2019deep}, electron paramagnetic resonance (EPR)~\cite{sunay2019small}, and positron annihilation spectroscopy (PAS)~\cite{von2012identification} have been commonly applied to capture point defect behavior. For example, DLTS measurements on n-type GaN identified the H1 trap level ($E_V + 0.87$~eV) as the (0/$-$) transition of substitutional carbon on nitrogen sites ($\mathrm{C_N}$)~\cite{kanegae2019deep}. EPR measurements following low-temperature electron irradiation revealed that Ga interstitials are mobile below room temperature~\cite{linde1997optical, bozdog1999optical, watkins2001radiation, chow2000detection}, and  EPR also revealed that the N interstitial prefers a split-interstitial configuration~\cite{von2012identification}. Because such measurements often cannot uniquely identify the defects responsible for the observed energy levels, integrating experimental observations with atomistic simulations is necessary to assign defect configurations to their energetics. 

Computational methodologies such as molecular dynamics (MD), which rely on an underlying force engine such as density functional theory (DFT) to compute atomic energies and forces, have become indispensable for investigating point defect energetics and defect–defect interactions~\cite{lyons2017computationally, gao2004intrinsic, gao2019point, hauck2024atomic, kyrtsos2016migration, hrytsak2021dft}. DFT provides predictions of defect formation energies and migration barriers across multiple charge states with high fidelity. However, this accuracy comes at a considerable computational expense, restricting simulations to relatively small supercells~\cite{lei2023comparative, lyons2017computationally, kyrtsos2016migration}, where spurious periodic image interactions can hinder proper energy convergence, as highlighted by Li et al.~\cite{li2020deep} and Hrytsak et al.~\cite{hrytsak2021dft}. Alternatively, MD simulations rely on parameterized interatomic potentials, such as Stillinger--Weber \cite{zhou2013relationship}, Tersoff \cite{nord2003molecular}, and more recently machine-learning potentials \cite{huang2023first}, to model defect energetics within significantly larger supercells. While MD with parameterized interatomic potentials generally sacrifices some energetic accuracy compared to DFT, it enables large-scale simulations and provides valuable insights into defect dynamics and structural evolution over extended spatial and temporal regimes~\cite{mahfuz2024microstructural, sequeira2021unravelling, sequeira2022examining}. Despite the availability of multiple interatomic potentials for GaN~\cite{zhou2013molecular, nord2003molecular, huang2023first}, existing force fields lack the capability to represent dynamic charge transfer and electrostatic coupling in the system for defect migration under external electric fields.

Reactive molecular dynamics simulations employing ReaxFF reactive force field provide a feasible pathway. ReaxFF enables large-scale atomistic simulations of defect dynamics while capturing dynamic bonding and partial charge redistribution that respond to the local chemical and electrostatic environment~\cite{senftle2016reaxff}. ReaxFF parameterizations have been established for diverse materials such as TiO$_2$~\cite{huygh2014development}, Si/SiO$_2$~\cite{nayir2019development}, MoS$_2$~\cite{ostadhossein2017reaxff}, and NiCr/FLiNaK~\cite{arkoub2024reactive, arkoub2025surface}, demonstrating their capability to capture point defect energetics, defect--defect interactions, and partial charge transfer phenomena. However, the application of such models to capture defect migration behavior in GaN under electrical fields remains unexplored.

In this work, we address this gap by developing a ReaxFF reactive force field for GaN. A comprehensive DFT-generated dataset is first employed to train and parameterize the GaN ReaxFF model. The resulting force field is then validated against key structural properties and defect energetics to ensure its reliability and transferability. Subsequently, the validated model is used to investigate the influence of an external electric field on point defect migration. We find that the electric field does not simply bias defect motion through a linear dipole--field coupling, but instead anisotropically changes migration barriers along the field direction. This behavior arises from concurrent partial-charge redistribution and local lattice distortion, both of which are captured natively within the reactive simulations, providing new mechanistic insights into charge-mediated defect behavior in GaN under realistic operating conditions.

\section{Computational Details}
\subsection{ReaxFF Force Field}
ReaxFF employs a dynamic bond-order approach that enables reactive simulations within a classical MD framework \cite{senftle2016reaxff,van2001reaxff}. Unlike traditional non-reactive force fields, ReaxFF allows bonds to form and break during the simulation and incorporates dynamic charge equilibration to capture partial charge transfer. The total potential energy is decomposed as
\begin{equation}
    \label{eqn: reaxff formalism}
    E_{total} = E_{bond} + E_{over} + E_{angle} + E_{tors} + E_{vdW} + E_{Coul} + E_{specific},
\end{equation}
with contributions from bond formation ($E_{bond}$), over-coordination penalty ($E_{over}$), three-body valence angles ($E_{angle}$), four-body torsions ($E_{tors}$), non-bonded van der Waals interactions ($E_{vdW}$), long-range electrostatics ($E_{Coul}$), and system-specific corrections such as lone-pair and conjugation terms ($E_{specific}$). Each contribution is modulated by the bond order BO$_{ij}$ between atoms $i$ and $j$, which evolves continuously with interatomic separation and the surrounding chemical environment. Full mathematical forms and physical bases are given by Senftle et al.~\cite{senftle2016reaxff} and van Duin et al.~\cite{van2001reaxff}. The force field parameters (global, element-specific, bond, off-diagonal, angle, and torsion terms) are optimized against first-principles reference data, as described next.

\subsection{First-Principles Calculations}

A comprehensive DFT reference dataset was generated using the Vienna Ab initio Simulation Package (VASP)~\cite{kresse1993initio, kresse1996efficiency, kresse1996efficient}. All calculations were spin-polarized. Core--valence interactions were treated within the projector-augmented wave (PAW) framework~\cite{blochl1994projector}, with Ga 3$d$ (Ga\_d potential) and N 2$s$ 2$p$ states included as valence electrons, and exchange--correlation effects were described by the Perdew--Burke--Ernzerhof (PBE) functional within the generalized gradient approximation~\cite{perdew1996generalized}. A plane-wave kinetic-energy cutoff of 400~eV was used, and Brillouin-zone sampling employed a fully converged $2\times2\times2$ \(\Gamma\)-centered k-point mesh. Supercell sizes and remaining settings are provided in the Supplementary Material (SM).

Equation-of-state (EOS) calculations were performed for both elemental Ga and GaN phases. For Ga, the training set included $\alpha$-Ga (orthorhombic), face-centered cubic (FCC), body-centered cubic (BCC), simple cubic, and tetragonal structures. For GaN, the training set included the wurtzite, rocksalt, and zincblende phases. In addition, the N$_2$ bond length and its corresponding energy profile were incorporated for an accurate description of molecular interactions.

Defect energetics formed a key component of the dataset. The training set included total energies of supercells containing neutral point defects, referenced to pristine supercells. The considered defects comprised vacancies ($V_\text{Ga}$ and $V_\text{N}$), antisites (N$_\text{Ga}$ and Ga$_\text{N}$), Frenkel pairs (Ga$_\text{FP}$ and N$_\text{FP}$), interstitial configurations (Ga$_{i1}$ at the octahedral site, Ga$_{i2}$ at the tetrahedral site, N$_{i1}$ at the octahedral site, and split interstitial N$_{i2}$), as well as the Schottky defect ($V_\text{Ga}$--$V_\text{N}$). The defect-structure energy ($E_{ds}$) was defined as
\begin{equation}
    \label{eqn: defect structure energy}
    E_{ds} = E_\text{defect} - E_\text{perfect}
\end{equation}
where $E_\text{defect}$ and $E_\text{perfect}$ are the total energies of the defective and pristine supercells, respectively.

\subsection{ReaxFF MD Simulations}
Following the parameterization of the force field, ReaxFF MD simulations were performed using LAMMPS~\cite{thompson2022lammps}, with structural visualization and analysis carried out in OVITO~\cite{stukowski2010visualization}. The ReaxFF MD calculations comprised two parts: (i) structural properties and (ii) defect migration barrier calculations under varying external electric fields. For structural optimization, a 2880-atom wurtzite supercell was relaxed under periodic boundary conditions, using the conjugate gradient algorithm with force and energy convergence criteria of $10^{-12}$~kcal/mol/Å and $10^{-12}$~kcal/mol, respectively. Charges were equilibrated at every step using the electronegativity equalization method (QEq)~\cite{mortier1985electronegativity}. The resulting system energies and atomic volumes were fitted to the third-order Birch--Murnaghan EOS~\cite{katsura2019simple}. Defect migration barriers were computed using the climbing-image nudged elastic band (CI-NEB) method~\cite{henkelman2000climbing, henkelman2000improved}, with energy and force convergence thresholds of $10^{-7}$~kcal/mol and $10^{-6}$~kcal/mol/Å, respectively. The cell was made non-periodic along the field direction and periodic in the remaining two directions. Charges were re-equilibrated at every step within the 2880-atom supercell. An electric field ranging from 0 to 0.04~V/Å (equivalent to 0--40~MV/cm), in increments of 0.01~V/Å, was applied along or perpendicular to the direction of defect migration pathway. This field range spans the GaN breakdown field (3.3--3.8~MV/cm, or 0.033--0.038~V/Å)~\cite{wong2021ultrawide, pearton2021radiation, maeda2021breakdown}. Atoms at the simulation-cell boundaries (3 {\AA} thickness) were held fixed to avoid rigid-body drift under the applied field.

\section{Results and Discussion}
\subsection{ReaxFF Training Data Comparison}
\begin{figure}[!h]
    \centering
    \begin{subcaptiongroup}
        \phantomcaption\label{fig:EOS wz GaN}
        \phantomcaption\label{fig:N2 bond}
        \phantomcaption\label{fig:EOS Ga a-c}
        \phantomcaption\label{fig:EOS Ga b}
    \end{subcaptiongroup}
    \includegraphics[width=1\linewidth]{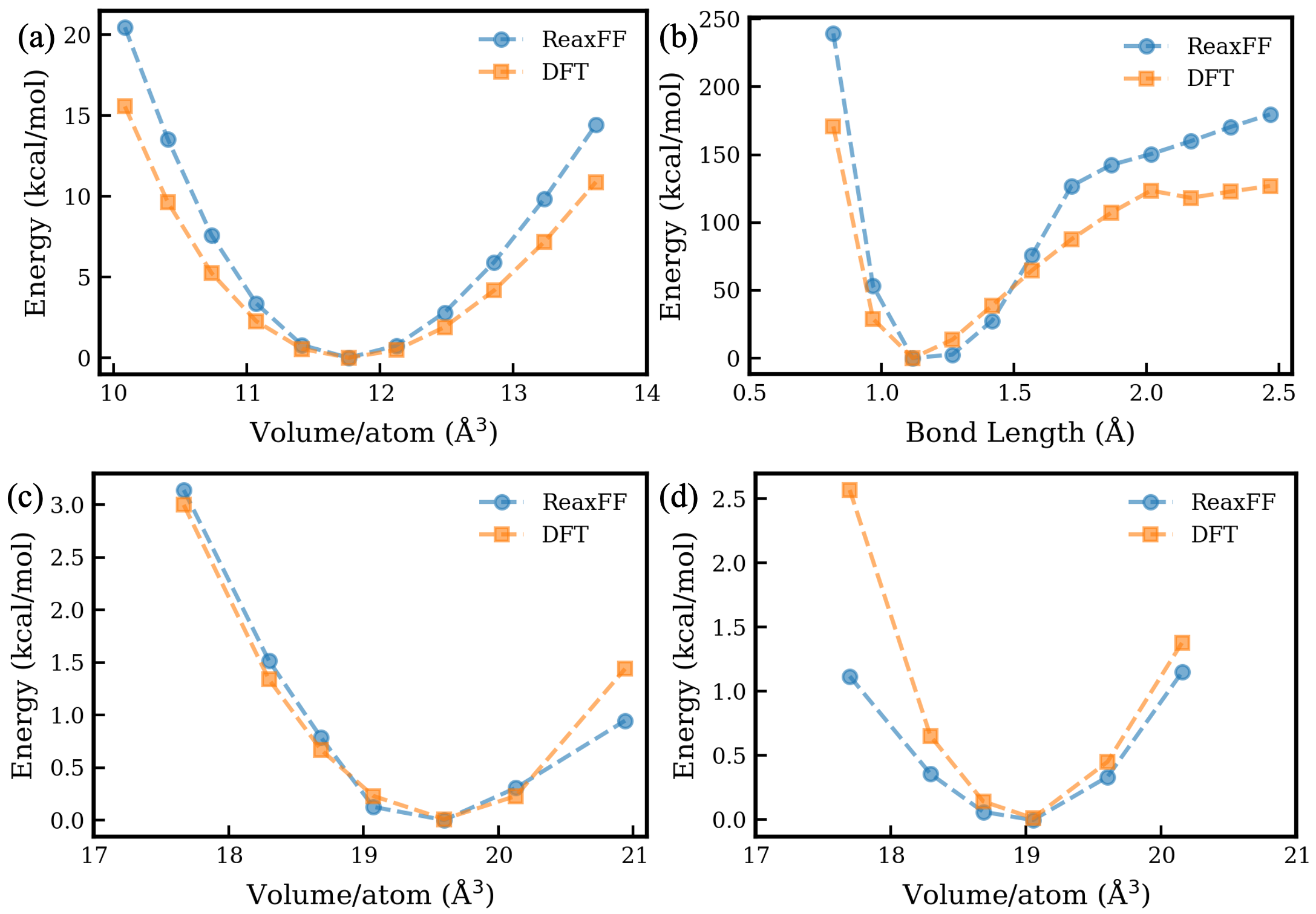}
    \caption{EOS data for wurtzite GaN \textbf{\subref{fig:EOS wz GaN}}, N$_2$ molecule bond length energy profile \textbf{\subref{fig:N2 bond}}, and EOS data for orthorhombic Ga where the volume changes along the $a$ and $c$ directions \textbf{\subref{fig:EOS Ga a-c}} or along the $b$ direction \textbf{\subref{fig:EOS Ga b}}. Blue circles and orange squares represent ReaxFF and DFT calculations, respectively.}
    \label{fig:EOS GaN N2}
\end{figure}

We first evaluated the EOS for each material included in the training dataset, encompassing multiple phases of GaN and Ga as well as the bond-length-dependent energy profile of the N$_2$ molecule. The EOS results for wurtzite GaN, orthorhombic Ga, and the N$_2$ molecule are presented in Fig.~\ref{fig:EOS GaN N2}; the remaining EOS data are provided in the SM.
For wurtzite GaN, the EOS data predicted by ReaxFF shows excellent agreement with the corresponding DFT calculations. Minor deviations are observed at extreme compressive and expansive volumes, where ReaxFF overestimates the energy relative to DFT (Fig.~\ref{fig:EOS wz GaN}). The energy--bond length curve of the N$_2$ molecule is also well reproduced near equilibrium, with ReaxFF predicting higher energies at bond lengths exceeding $\sim$1.5~Å (well beyond equilibrium). For orthorhombic Ga, two separate EOS calculations are shown in Figs.~\ref{fig:EOS Ga a-c}--\subref{fig:EOS Ga b}: one corresponding to volumetric changes along the $a$--$c$ directions and another along the $b$ direction. In both cases, the ReaxFF results closely follow the DFT trends, demonstrating that the model captures the structural energetics across different bonding environments.  

The equilibrium structural properties of wurtzite GaN were extracted from the EOS fit and are summarized in Table~\ref{table: eq prop data}. The equilibrium energy and volume agree closely with prior theoretical values~\cite{pandey1994ab}. The bulk modulus, $B = 169$~GPa, falls slightly below the calculated DFT measurements of 172~GPa. Lattice parameters were obtained by fully relaxing the structure through energy minimization, and the predicted values closely agree with DFT results calculated using the PBE exchange--correlation functional~\cite{gonzalez2014comparative}.  

\begin{table}[h!]
\centering
\caption{Comparison of calculated and literature values of equilibrium properties for wurtzite GaN.}
\label{table: eq prop data}
\begin{tabular}{|c|c|c|}
\hline
\textbf{Parameter} & \textbf{This work} & \textbf{Literature/DFT} \\
\hline
 Minimum Energy, $E_0$ (eV) &  $-6.08$ & $-6.08$~\cite{pandey1994ab} \\
 Minimum Volume, $V_0$ (Å$^3$)  & $11.77$ & $11.94$~\cite{pandey1994ab} \\
 Bulk Modulus, B (GPa) &  $169.75$ & $172$\\
 \multirow{2}{*}{Lattice constants (Å)}
  & $a = 3.21$ & $a = 3.22$~\cite{gonzalez2014comparative} \\
  & $c = 5.25$ & $c = 5.24$~\cite{gonzalez2014comparative}\\
\hline
\end{tabular}
\end{table}

\begin{figure}[!h]
    \centering
    \begin{subcaptiongroup}
        \phantomcaption\label{fig:defect dE}
        \phantomcaption\label{fig:defect E_form}
    \end{subcaptiongroup}
    \includegraphics[width=1\linewidth]{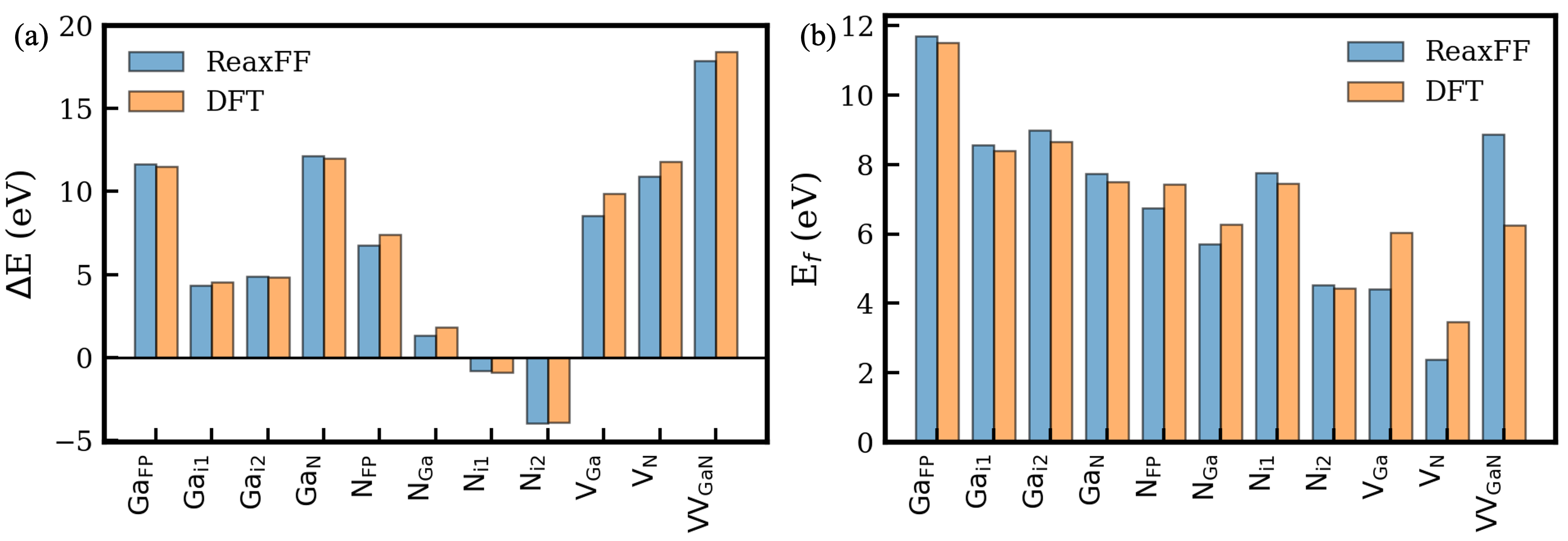}
    \caption{(a) Energy difference between defect structure and perfect structure. (b) Defect formation energies under N-rich conditions. Blue and orange bars correspond to ReaxFF and DFT calculations, respectively.}
    \label{fig:defect energy}
\end{figure}

We further assessed the performance of the force field by comparing defect energetics against corresponding DFT calculations. Fig.~\ref{fig:defect dE} shows the energy differences between defective and pristine supercells, computed using ReaxFF and DFT calculated using both ReaxFF and DFT for the same initial supercell configurations. The maximum deviation in structure energy relative to DFT is 26\% across the defect configurations considered (SM, Table~S1-S2). In addition, defect formation energies under nitrogen-rich conditions were calculated for neutral defect types examined in this study. These values were similarly compared with DFT predictions (Fig.~\ref{fig:defect E_form}). The comparatively larger energy variations are observed for the vacancies. Overall, those validation calculations demonstrate the capability of the developed force field to reasonably capture defect energetics in GaN.

\subsection{Defect Migration under External Electric Field}
\begin{figure}[!h]
    \centering
    \includegraphics[width=1.0\linewidth]{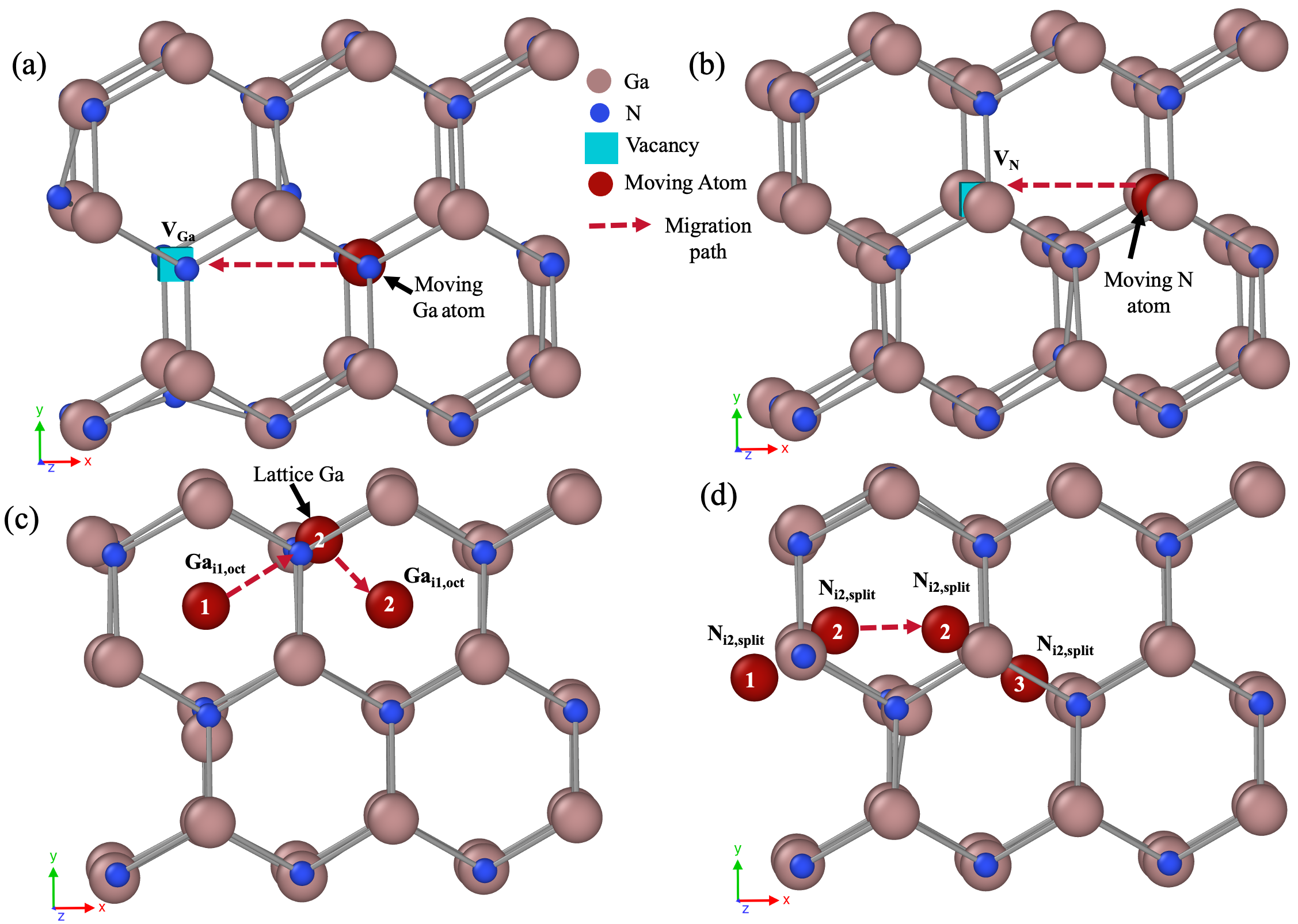}
    \caption{Migration directions for (a) V$_\text{Ga}$, (b) V$_\text{N}$, (c) Ga$_\text{i1,oct}$, and (d) N$_\text{i2,split}$, respectively.}
    \label{fig:migration path}
\end{figure}

Prior theoretical investigations have examined the migration barriers of intrinsic point defects in GaN using both first-principles and empirical approaches under zero field~\cite{kyrtsos2016migration, limpijumnong2004diffusivity, zhu2023computational, hrytsak2021dft, li2024deep, reza2026evaluation}. Hence, for benchmarking purposes, we first consider the migration barriers of intrinsic point defects. For vacancy diffusion, two distinct migration pathways are considered: an in-plane path, where the vacancy moves within the basal plane perpendicular to the $c$-axis, and an out-of-plane path, where migration occurs along the direction parallel to the $c$-axis. Interstitials migrate through two primary mechanisms: the $c$-channel pathway and the interstitialcy mechanism. In the $c$-channel process, the interstitial atom diffuses along the open hexagonal channels parallel to the $c$-axis of the wurtzite structure. The interstitialcy mechanism, however, involves a concerted atomic exchange, whereby the interstitial displaces a lattice atom from its site and occupies that position, effectively converting the displaced atom into a new interstitial. Previous computational studies have consistently shown that in-plane vacancy migration and the interstitialcy pathway represent the energetically preferred diffusion mechanisms in GaN~\cite{kyrtsos2016migration, limpijumnong2004diffusivity, zhu2023computational, hrytsak2021dft, li2024deep, reza2026evaluation}. Guided by these findings, we focus on the energetically favorable migration pathways of point defects as depicted in Fig.~\ref{fig:migration path}, and compute the corresponding migration barriers using ReaxFF. It is important to note that vacancy migration requires movement of only one atom (Fig.~\ref{fig:migration path}(a-b)) while interstitial migration requires multiple atom movements. Ga interstitialcy mechanism involves two atoms (denoted by 1 and 2) in Fig.~\ref{fig:migration path}(c) and N interstitialcy mechanism involves three-atom movement (denoted by 1, 2, and 3) in Fig.~\ref{fig:migration path}(d). 

\begin{table}[h!]
\centering
\caption{Neutral point defect migration energies for vacancies and interstitials in wurtzite GaN. Alternate charge states from the literature are indicated in parentheses.}
\label{table: em data}
\begin{tabular}{|c|c|c|}
\hline
\textbf{Defect Type} & \textbf{This work (eV)} & \textbf{Literature (eV)} \\
\hline
\multirow{3}{*}{$V_{\text{Ga}}$}
 & \multirow{3}{*}{$2.35$}  & $1.90$~\cite{warnick2011room} \\
 &  & $2.38$~\cite{hrytsak2021dft} \\
 &       & $2.50$~\cite{kyrtsos2016migration} \\
 \hline
 \multirow{3}{*}{$V_{\text{N}}$}
  & \multirow{3}{*}{$2.17$} & $2.0$~\cite{warnick2011room}  \\
  & & $2.7$~\cite{kyrtsos2016migration} \\
  & & $3.13$~\cite{hrytsak2021dft} \\
  \hline
\multirow{2}{*}{Ga$_{i1,\mathrm{oct}}$} &  \multirow{2}{*}{$0.88$} & $0.7$ ($+3$)~\cite{kyrtsos2016migration}  \\
 & & $0.9$ ($+3$)~\cite{limpijumnong2004diffusivity} \\
 \hline
\multirow{2}{*}{N$_{i2,\mathrm{split}}$} & \multirow{2}{*}{$1.03$} & $1.4$~\cite{zhu2023computational} \\
 & & $2.4$~\cite{kyrtsos2016migration} \\
\hline
\end{tabular}
\end{table}


Table~\ref{table: em data} summarizes the migration barriers. The in-plane migration barrier predicted by ReaxFF for neutral V$_\text{Ga}$ is $2.35$ eV, which lies within the reported literature range of $1.90$ eV to $2.50$ eV and matches the most recent DFT value of 2.38~eV~\cite{hrytsak2021dft}. The neutral V$_\text{N}$ barrier is 2.17~eV, near the lower end of the reported range of 2.0--3.13~eV. 
For the octahedral Ga interstitial (Ga$_\text{i1,oct}$),  published computational data pertain to the $+3$ charge state (0.7 and 0.9~eV~\cite{kyrtsos2016migration, limpijumnong2004diffusivity}).The ReaxFF value of 0.88~eV falls within the reported $+3$ range, suggesting that the neutral and $+3$ barriers are similar in magnitude in this system. For the split nitrogen interstitial (N$_\text{i2,split}$), ReaxFF predicts 1.04~eV, while literature values show substantial variation, ranging from 1.4 to 2.4~eV~\cite{zhu2023computational, kyrtsos2016migration}. The field-dependent trends discussed below are based on relative changes in the barrier, so the systematic offset in the absolute barrier is expected to have limited impact on the resulting trends.

We next examine how an external electric field modifies these barriers. All energy values in this subsection are referenced to the initial minimized system energy at zero field. As a comparison baseline, we adopt the linear dipole--field approximation introduced by El-Sayed et al.~\cite{el2018effect},
\begin{equation}
    \label{eqn: efield barrier change}
    \Delta E_B^F = -\mu_{\mathrm{eff}} \cdot E,
\end{equation}
where $\Delta E_B^F$ is the field-induced change in the migration barrier, $\mu_{\mathrm{eff}}$ is the effective dipole moment associated with the migrating defect, and $E$ is the applied field. Departures from this linear expression indicate higher-order contributions from partial-charge redistribution and lattice distortion not captured by a fixed-dipole model.

\begin{figure}[!h]
    \centering
    \begin{subcaptiongroup}
        \phantomcaption\label{fig:vga E_x path}
        \phantomcaption\label{fig:vga E_x barrier}
        \phantomcaption\label{fig:vn E_x path}
        \phantomcaption\label{fig:vn E_x barrier}
    \end{subcaptiongroup}
    \includegraphics[width=1.0\linewidth]{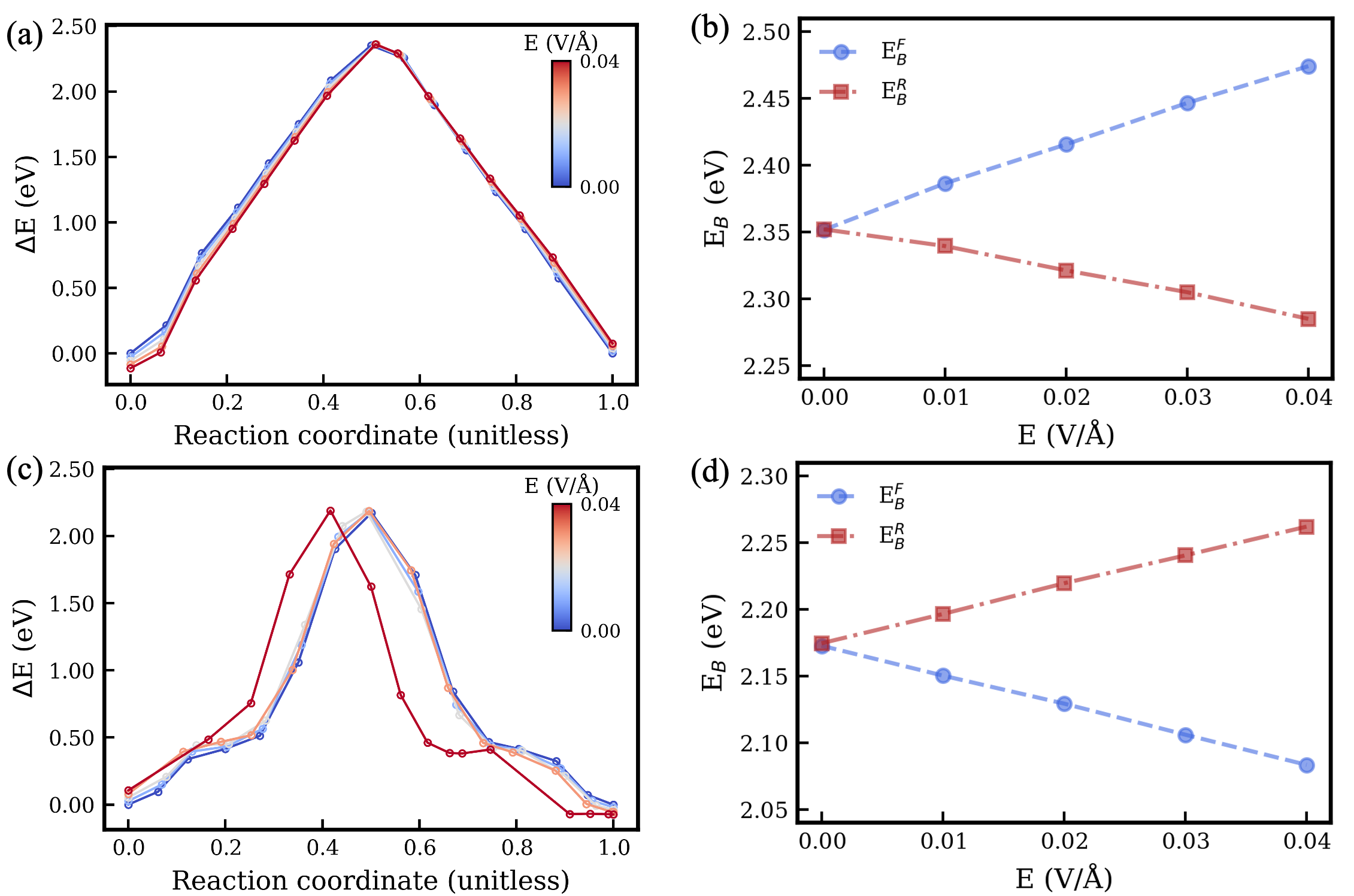}
    \caption{Effect of an external electric field on the total system energy relative to zero field energy at zero reaction coordinate and defect migration barrier for V$_\text{Ga}$ (a-b) and V$_\text{N}$ (c-d). The electric field is applied along the $x$-direction (parallel to the migration path). For the field-dependent migration barriers, $E_B^F$ values are given by blue circles while $E_B^R$ values are shown as red squares.
    }
    \label{fig:vga+vn efield}
\end{figure}

\begin{figure}[!h]
   \centering
   \begin{subcaptiongroup}
       \phantomcaption\label{fig:vga charge}
       \phantomcaption\label{fig:vn charge}
   \end{subcaptiongroup}
   \includegraphics[width=1.0\linewidth]{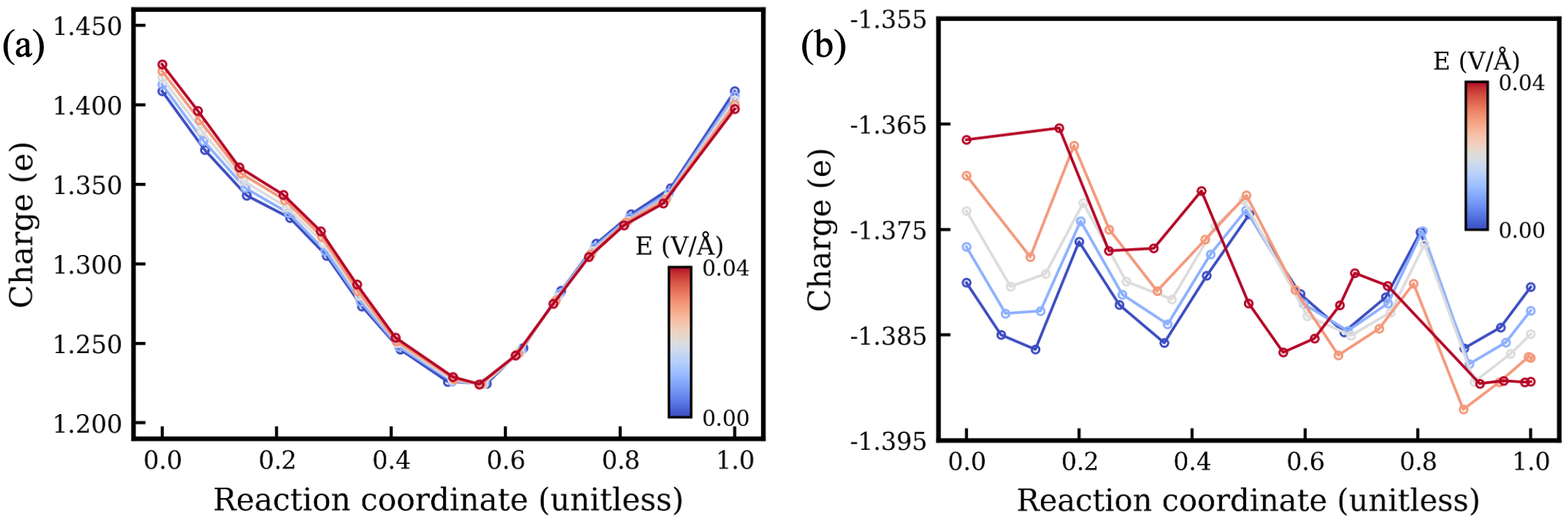}
   \caption{Charge distribution variation with applied electric field along $x$-direction for the moving atom in V$_\text{Ga}$~\subref{fig:vga charge} and V$_\text{N}$~\subref{fig:vn charge}.}
   \label{fig:vacancy charge main}
\end{figure}
The influence of an external electric field on the migration of V$_\text{Ga}$ is illustrated in Fig.~\ref{fig:vga+vn efield}(a--b). With the field parallel to the vacancy migration path ($x$-direction), the migrating Ga atom, which carries a positive partial charge, moves antiparallel to the applied field (Fig.~\ref{fig:migration path}(a)).  The forward ($E_B^F$) and reverse ($E_B^R$) migration barriers, extracted from the energy profiles in Fig.~\ref{fig:vga E_x path}, are plotted against field strength in Fig.~\ref{fig:vga E_x barrier}. With increasing field, $E_B^F$ rises approximately linearly while $E_B^R$ decreases approximately linearly, but with unequal slopes.  
Notably, the magnitudes of these changes are asymmetric: as the field increases from 0 to 0.04~V/Å, $E_B^F$ rises by approximately 0.15~eV while $E_B^R$ decreases by only about 0.10~eV. This non-equivalent response indicates that the field-induced barrier modification does not follow the linear relationship predicted by Eq.~\ref{eqn: efield barrier change}, which assumes a configuration-independent effective dipole and produces equal-magnitude, opposite-sign changes in $E_B^F$ and $E_B^R$ under an applied field. The deviation arises primarily from two factors. First, the relatively larger atomic size of Ga induces local lattice strain and structural distortion during migration.  Second, the partial charge on the migrating Ga atom varies along the reaction coordinate, and the applied field further redistributes and breaks the symmetry of this distribution (Fig.~\ref{fig:vga charge}). In contrast, when the field is applied perpendicular to the migration direction (along $z$), the energy landscape is largely unchanged (Fig.~S3(a), SM) and the barriers exhibit negligible variation (Fig.~S3(b), SM); hence, the response is governed by the field component projected onto the reaction coordinate.

The $V_\text{N}$ responds differently from $V_\text{Ga}$, owing primarily to its smaller atomic size and the correspondingly reduced lattice strain during migration. When the electric field is applied parallel to the defect migration pathway, the total relative energy profile shifts along the reaction coordinate, with the effect becoming most pronounced at field strengths around $0.04$ V/Å (Fig.~\ref{fig:vn E_x path}). Because the strain contribution and the charge variation of the migrating N atoms along the path are minor (Fig.~\ref{fig:vn charge}), $E_B^F$ and  $E_B^R$  extracted from these profiles vary linearly with field strength across the full range and shift by nearly equal magnitudes (Fig.~\ref{fig:vn E_x barrier}), in close agreement with the linear El-Sayed picture of Eq.~\ref{eqn: efield barrier change}. The sign of the barrier modification is opposite to that of $V_\text{Ga}$, reflecting the negative partial charge of the migrating nitrogen atom (1.41e for Ga and -1.38e for N). When the electric field is oriented perpendicular to the migration direction, both $E_B^F$ and $E_B^R$  decrease slightly with increasing field strength (Fig.~S3(c-d), SM). This modest reduction is attributed to subtle field-induced changes in the partial charge distribution, and indicates that the charge-redistribution mechanism remains active even without geometric alignment, distinguishing $V_\text{N}$ from $V_\text{Ga}$.

\begin{figure}[!h]
    \centering
    \begin{subcaptiongroup}
        \phantomcaption\label{fig:gai E_x path}
        \phantomcaption\label{fig:gai E_x barrier}
        \phantomcaption\label{fig:ni E_x path}
        \phantomcaption\label{fig:ni E_x barrier}
    \end{subcaptiongroup}
    \includegraphics[width=1.0\linewidth]{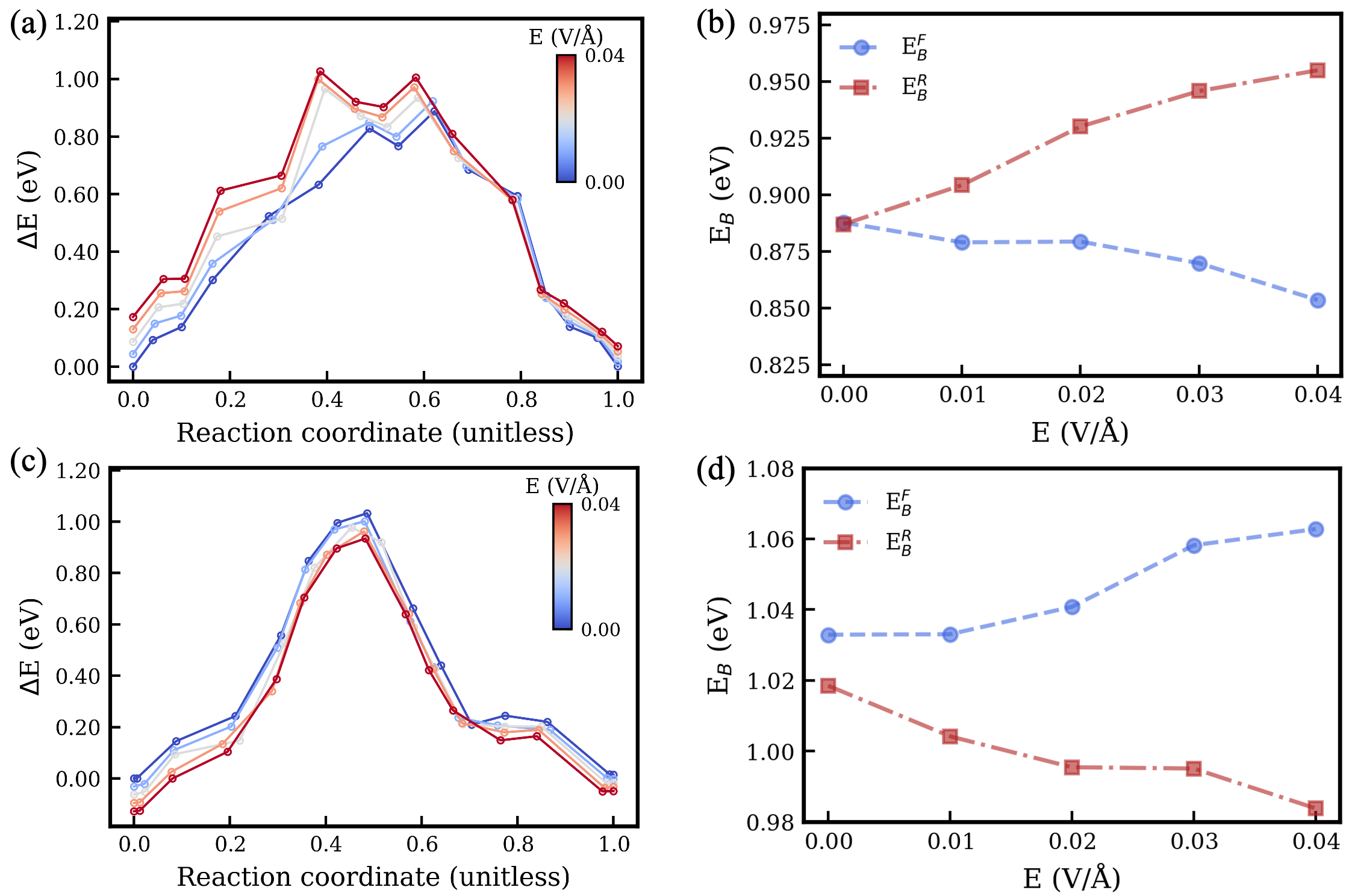}
    \caption{ Effect of an external electric field on the total system energy relative to zero field energy at zero reaction coordinate and defect migration barrier for Ga$_\text{i1,oct}$ (a-b) and N$_\text{i2,split}$ (c-d). The electric field is applied along the $x$-direction (parallel to the migration path). For the field-dependent migration barriers, $E_B^F$ values are given by blue circles while $E_B^R$ values are shown as red squares.}
    \label{fig:gai+ni efield}
\end{figure}

\begin{figure}[!h]
   \centering
   \begin{subcaptiongroup}
       \phantomcaption\label{fig:gai charge}
       \phantomcaption\label{fig:ni charge}
   \end{subcaptiongroup}
   \includegraphics[width=1.0\linewidth]{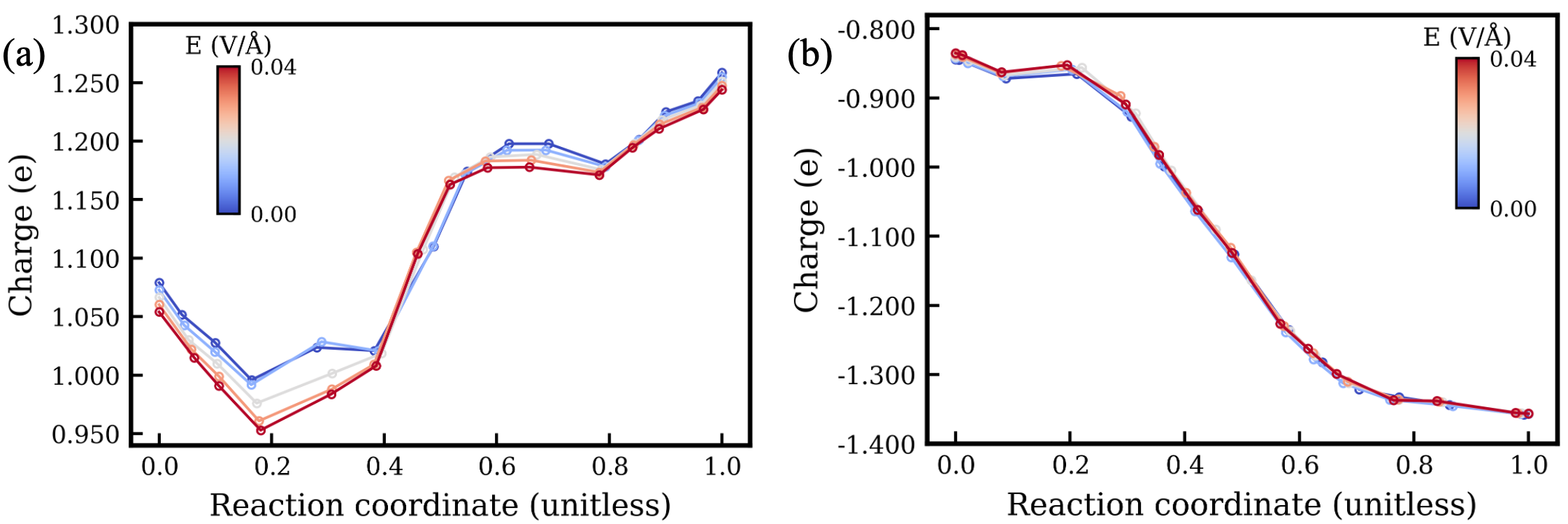}
   \caption{Charge distribution variation with applied electric field along $x$-direction for the moving atom in Ga$_\text{i1,oct}$~\subref{fig:gai charge} and N$_\text{i2,split}$~\subref{fig:ni charge}. The charge distribution is shown for atom 1 during Ga$_\text{i1,oct}$ migration and for atom 1 during N$_\text{i2,split}$ migration, as defined in Fig.~\ref{fig:migration path}. Additional charge distributions are provided in the SM.}
   \label{fig:interstitial charge main}
\end{figure}

Next, the influence of an external electric field on the migration of Ga$_{i1,\mathrm{oct}}$ is shown in Fig.~\ref{fig:gai+ni efield}(a--b). With the field applied along the $x$-direction, the energy profile differs qualitatively from the vacancy cases: instead of a single peak at the saddle, the interstitialcy pathway produces a profile with two peaks in the transition region (Fig.~\ref{fig:gai E_x path}). The relative heights of these peaks depend on the field: the second peak dominates at low fields, while the first peak dominates at high fields. This crossover indicates a field-induced shift in the rate-limiting saddle-point configurations. The forward and reverse barriers are defined as the difference between the highest point along the path and the initial or final state energy, respectively. In contrast to the vacancy cases, these barriers do not vary linearly with field strength, and the reverse barrier $E_B^R$ exhibits a larger overall change than the forward barrier $E_B^F$ (Fig.~\ref{fig:gai E_x barrier}). This is attributed to two factors. First, the migration pathway has a non-negligible component along the $y$ and $z$-direction and the intermediate configurations vary strongly with the field strength. Second, the interstitialcy mechanism involves cooperative motion of two atoms (a lattice host atom and the interstitial), producing a more complex partial-charge redistribution and strain response. The variability of charge can be seen from the charge distribution of the interstitial atom (number 1 indicated in Fig. ~\ref{fig:migration path}), as shown in Fig.~\ref{fig:gai charge}; other distributions are shown in Fig.~S6 in SM). With the field applied along $z$, both barriers show a nonlinear trend within a range of approximately 0.85--0.90~eV (Fig. S4(b), SM), indicating modest sensitivity to the perpendicular field  (stronger than vacancy cases).

Finally, we analyze the influence of an external electric field on the migration of  N$_{i2,\mathrm{split}}$. The energy profile exhibits a single dominant saddle point together with a weakly metastable state near the end of the migration pathway (Fig.~\ref{fig:ni E_x path}). Note that the initial and final configurations along this pathway correspond to two inequivalent split-interstitial orientations, consistent with the migration mechanism reported by He et al.~\cite{he2020dynamics}, so $E_B^F$ and $E_B^R$ are slightly different even under zero field. Both barriers respond nonlinearly to the applied field, with the forward barrier rising and the reverse barrier falling (Fig.~\ref{fig:ni E_x barrier}) in unequal slopes: $E_B^R$ exhibits a larger change than $E_B^F$. This is due to i) a small $y$ and $z$-component in the migration pathway that produces partial misalignment with the applied field, and ii) the interstitialcy mechanism, which here involves cooperative motion of three atoms and yields a more complex partial-charge (Charge distribution of atom 1 represented in Fig.~\ref{fig:migration path}(d) is shown in Fig.~\ref{fig:ni charge}; other atom charge distributions are shown in Fig.~S7 in SM) and local-strain response. With the field applied along $z$, the migration is predominantly along $x$, so the perpendicular field couples only weakly to the reaction coordinate; the resulting barrier changes are minimal compared to Ga interstitial (Fig.~S4(c--d) in SM).

\section{Discussion}
The four defects examined here reveal a defect-specific pattern of field response that cannot be captured by a single linear coefficient. Vacancies, whose migration involves only the displaced atom, show responses that are either nearly symmetric (V$_\mathrm{N}$) or mildly asymmetric (V$_\mathrm{Ga}$). Interstitials, whose migration proceeds via cooperative motion of two or three atoms, show pronounced nonlinearity and a forward/reverse slope mismatch in which $|\Delta E_B^R| > |\Delta E_B^F|$ for both Ga$_{i1,\mathrm{oct}}$ and N$_{i2,\mathrm{split}}$. The directional asymmetry also implies defect-specific net migration tendencies under field: V$_\mathrm{N}$ and Ga$_{i1,\mathrm{oct}}$ are driven in one direction by the reduction of their forward barriers, while V$_\mathrm{Ga}$ and N$_{i2,\mathrm{split}}$ are driven in the opposite direction by the reduction of their reverse barriers. The sign of the response correlates with the sign of the partial charge on the migrating species, but its magnitude and the degree of nonlinearity reflect the defect's specific bonding and migration geometry. Analogous field-induced asymmetry has been reported in other partially ionic semiconductors such as MgO, where O interstitials and vacancies show qualitatively similar behavior~\cite{el2018effect}.

The departure from a linear field--barrier relationship stems from the implicit rigid-dipole assumption underlying the El-Sayed expression of Eq.~\ref{eqn: efield barrier change} and the related expressions used by Warnick et al.~\cite{warnick2011room} and Strand et al.~\cite{strand2020effect}. In that picture, the defect is treated as a point dipole with a fixed effective moment that translates rigidly along the reaction coordinate, so the field couples linearly to position. The configurations sampled along an NEB path, however, differ not only by the position of the migrating atom but also by the local lattice strain around it and the partial charges it carries, all of which evolve continuously between the initial state, the saddle, and the final state. As a result, the effective dipole varies along the path. For defects in which these variations are modest (V$_\mathrm{N}$), the constant-dipole picture remains a reasonable approximation, and our results recover the linear, symmetric forward/reverse response that the rigid-dipole model predicts. For defects in which lattice strain and partial-charge variation are large (V$_\mathrm{Ga}$), or in which cooperative atomic motions add additional coupling (Ga$_{i1,\mathrm{oct}}$, N$_{i2,\mathrm{split}}$), the deviation shows up; most noticeably for Ga$_{i1,\mathrm{oct}}$, where the field shifts the energy profile sufficiently to switch the rate-limiting saddle point between low and high fields.

The quantitative kinetic impact of these field-induced barrier changes depends on the absolute barrier height. The Arrhenius factor $D \propto \exp(-E_B / k_B T)$ suggests a barrier reduction $\Delta E_B$ by $\exp(-\Delta E_B / k_B T)$, which at room temperature ($k_B T \approx 0.025$~eV) corresponds to roughly an order of magnitude for every 0.06~eV. In prior work, native-defect migration in oxides has been shown to become appreciable at room temperature once barriers fall below approximately 1~eV~\cite{janotti2007native, iddir2007diffusion, capron2007migration}. For the interstitial defects examined here (Ga$_{i1,\mathrm{oct}}$ and N$_{i2,\mathrm{split}}$, barriers near 1~eV), the maximum field-induced reductions of $\sim 0.05$~eV at 0.04~V/Å therefore yield rate enhancements of an order of magnitude, sufficient to shift these defects mobility at typical device operating temperatures, particularly when combined with the elevated junction temperatures characteristic of HEMT operation~\cite{warnick2011room}. Such field--defect coupling is expected to be stronger for highly charged defects from carrier trapping, and in regions of locally strong field (heterointerfaces, gate edges, breakdown regions) \cite{meneghesso2008reliability, tapajna2010integrated}. By comparison, for the vacancy defects (V$_\mathrm{Ga}$ and V$_\mathrm{N}$, barriers above 2~eV), the field-induced reductions leave the absolute barrier well above the threshold for thermally activated diffusion on device-operation timescales.

These observations carry important implications for the operating behavior and long-term reliability of GaN-based devices. Modern AlGaN/GaN HEMTs and vertical power devices routinely sustain channel-region fields of 1--3~MV/cm under bias, with peak fields at the gate edge approaching the breakdown limit~\cite{meneghesso2008reliability, tapajna2010integrated}, which overlaps the range explored here (0--4~MV/cm). These external fields coexist with the large intrinsic polarization fields at AlGaN/GaN heterointerfaces~\cite{ambacher1999two}. Two consequences are expected. First, defects whose neutral migration barriers approach the room-temperature mobility threshold of $\sim$1~eV (Ga$_{i1,\mathrm{oct}}$ and N$_{i2,\mathrm{split}}$ in this study) become substantially more mobile in the high-field regions of operating devices, contributing to the trap generation, threshold-voltage drift, and gate-leakage instabilities that have long limited GaN HEMT lifetime. Second, for the mobile species, the directional asymmetry favors net drift toward specific regions of the device; accumulated over operational timescales, this may drive long-range defect redistribution and concentration buildup at structural features such as interfaces and gate edges. These findings are most relevant to the radiation-tolerance applications involving wide-bandgap nitride electronics~\cite{pearton2015ionizing, mahfuz2024microstructural, jin2025examining}, where the field environment itself acts as a continuous driving force for defect redistribution and evolution.

Finally, we acknowledge the limitations of the present method, which motivates future study. The QEq scheme~\cite{mortier1985electronegativity} used by ReaxFF~\cite{senftle2016reaxff} distributes electronic charge among atoms based on instantaneous geometry and electronegativity equalization; it provides a classical-electrostatics treatment that does not resolve discrete charge states as commonly seen in DFT calculations. Given the charge balance, our calculations therefore correspond to neutral defects in the DFT sense. The neutral-defect results presented here therefore capture the qualitative coupling between field, lattice, and charge distribution, but do not provide an explicit treatment of charge-state-dependent migration. Furthermore, real defects in operating devices frequently occupy high charge states (e.g., $V_\mathrm{Ga}^{3-}$ in n-type GaN), which carry larger absolute charges and likely larger effective dipoles; the current results therefore represent a lower bound on the field response of charged defects under operational conditions. Under sufficiently strong fields, defects may also undergo charge-state transitions~\cite{wolfowicz2018electrometry} (e.g., $V_\mathrm{Ga}^{0/-/2-/3-}$) whose detailed kinetics fall outside the QEq description. Despite those, it is expected that the findings on the nonlinear, defect-specific character of field--defect coupling arising from partial-charge redistribution and lattice strain along the migration pathway should still survive more sophisticated electronic treatments to relax the QEq approximations~\cite{vondrak2025pushing}. 

\section{Conclusion}
We have developed and validated a ReaxFF reactive force field for GaN against a comprehensive DFT training set spanning structural properties and point-defect energetics, and used it to investigate the response of point defects to externally applied electric fields. For vacancies, each barrier responds linearly to a field applied along the migration path; the forward and reverse changes are mildly asymmetric in $V_\mathrm{Ga}$ and nearly symmetric in $V_\mathrm{N}$. For interstitials, both Ga$_{i1,\mathrm{oct}}$ and N$_{i2,\mathrm{split}}$ exhibit pronounced nonlinear and asymmetric responses, including, for Ga$_{i1,\mathrm{oct}}$, a field-induced shift in the rate-limiting saddle-point configuration. Perpendicular fields, as expected, produce minor barrier changes in all four cases due to lattice relaxation. These results establish that field--defect coupling in GaN cannot be captured by a single linear dipole--field coefficient: the effective dipole varies continuously along the migration pathway, and the resulting response is anisotropic and defect-specific. The kinetic significance is greatest for the interstitial defects, whose intrinsic barriers approach the room-temperature mobility threshold of $\sim$1~eV. These effects impact directly on the long-term reliability of GaN-based electronics in radiation environments.

\section{Supplementary Material}
The Supplementary Material provides additional information, as referenced in the manuscript.

\section{Acknowledgments}
This work was supported by the Air Force Office of Scientific Research under Award No. FA9550-22-1-0308. Computations for this research were performed on the Pennsylvania State University’s Institute for Computational and Data Sciences’ Roar Collab Cluster.

\section{Data Availability}
All relevant data are available within the article and its accompanying Supplementary Materials.

\bibliography{references}

\end{document}